\documentclass[useAMS,usenatbib,twocolumn]{aastex62}
\usepackage{times,graphicx,amsmath,amsfonts,amssymb}
\usepackage{epsfig}
\usepackage{enumitem}
\usepackage{tabularx}
\usepackage{booktabs}
\usepackage{multirow}

\usepackage{array}
\newcolumntype{L}[1]{>{\raggedright\let\newline\\\arraybackslash\hspace{0pt}}m{#1}}
\newcolumntype{C}[1]{>{\centering\let\newline\\\arraybackslash\hspace{0pt}}m{#1}}
\newcolumntype{R}[1]{>{\raggedleft\let\newline\\\arraybackslash\hspace{0pt}}m{#1}}

\newcommand{\hi}{\mbox{H{\scriptsize I}}}

\def\ltsim{\lower.5ex\hbox{$\; \buildrel < \over \sim \;$}}
\def\gtsim{\lower.5ex\hbox{$\; \buildrel > \over \sim \;$}}
\def\ltsim{\lower.5ex\hbox{$\; \buildrel < \over \sim \;$}}
\def\gtsim{\lower.5ex\hbox{$\; \buildrel > \over \sim \;$}}

\def\be{\begin{equation}}
\def\ee{\end{equation}}
\def\ba{\begin{eqnarray}}
\def\ea{\end{eqnarray}} 

\mathchardef\mhyphen="2D 

\def\kms{\, {\rm km }\, {\rm s}^{-1}}
\def\rtr{r_\mathrm{tr}}

\def\rv{r_\textrm{v}}

\def\Mv{M_\textrm{h}}
\def\Ms{M_*}
\def\Mg{M_\textrm{g}}
\def\fg{f_\textrm{g}}

\def\rhoobs{\rho^\textrm{E}_{_*}}
\def\mnras{MNRAS}

\def\dd{\mathrm{d}}

\def\vc{V_\mathrm{c}}

\newcommand{\NDF}{\textsc{[KKS 2000]04}}

\shorttitle{Diffuse with little dark matter}
\shortauthors{A. Nusser}
\begin{document}

\title{A scenario for ultra-diffuse satellite galaxies  with low velocity dispersions: the case of \NDF }
\email{adi@physics.technion.ac.il} 
\author{Adi Nusser}
\affiliation{Department of Physics and the Asher Space Research Institute, Israel 
Institute of Technology Technion, Haifa 32000, Israel\\ and \\
Guangdong Technion-Israel Institute of Technology, Shantou 515063, P.R.China }


\begin{abstract}

A scenario for achieving a low velocity dispersion  for the galaxy  \NDF\ (aka NGC 1052-DF2) and similar galaxies is presented.
 A progenitor halo corresponding to a $z=0$ halo of mass $\ltsim 5\times 10^{10}\; \textrm{M}_\odot$ and a low concentration parameter (but consistent with cosmological simulations)  infalls onto a Milky Way-size  host at early times.  {Substantial removal of cold gas}  from the inner regions by supernova feedback and ram pressure, assisted by tidal stripping of the dark matter in the outer regions,  leads to a substantial  reduction of the velocity dispersion of stars within one effective radius.
In this framework, the observed stellar content of \NDF\ is associated with a progenitor mass close to that inferred from the global stellar-to-halo-mass ratio.  As far as the implications of kinematics are concerned, even if at a $\sim 20 $ Mpc distance, it is argued that  \NDF\  is  no more  peculiar  than  numerous early type galaxies with seemingly little total dark-matter content.

\end{abstract}

\keywords{galaxies: halos - cosmology: theory, dark matter, galaxies}

\section{Introduction}
\label{sec:intro}
Dark matter (DM) is subdominant within the effective radii (enclosing  half the total light) $R_\textrm{e}$ of ordinary  early type galaxies \citep[ETG; see][for a review]{Cappellari2016}. The inferred average  DM fraction within $R_\textrm{e}$ is $  \sim 30\%$,  and a fraction of these galaxies remains  consistent without any  DM in the inner regions. These findings are obtained using analyses of the stellar kinematics as 
well as gravitational lensing \citep[e.g.][]{Treu2004,Mamon2005a,Auger2009,Thomas2011}. 
Outer parts of ETGs  extending to several $R_\textrm{e}$ could be directly probed with the kinematical measurements of planetary nebulae (PN) that are detectable through strong emission lines. While it is reasonable that the stellar kinematics in the inner regions is only mildly  dominated  by DM, 
a puzzle emerged when galaxies with  PN kinematics  required very little DM also in the outer regions \citep{Romanowsky2003}.
However, a high DM content  has been demonstrated to reproduce the observed low velocity dispersion of PNs in a realistic scenario for the formation of ETGs, where the PNs are expected to follow elongated orbits \citep{2005Natur.437..707D}. 

The global  stellar-to-halo-mass ratio \citep[SHMR; ][]{Behroozi2010,Moster2013,Rodriguez-Puebla2017} can be used to derive  the peak \footnote{The peak halo mass is the maximum mass the halo acquires 
throughout its history, and it can be quite different from its current mass. }  viral mass of the halo, $\Mv$, from its observed stellar mass, $\Ms$.
Assuming DM halos with an NFW density profile \citep{Navarro1996},  the virial mass  tuned   to match the stellar kinematics of SDSS ellipticals yields   $\Mv /\Ms\sim 3$ (with a large scatter) in galaxies with $\Ms\sim 5 \times 10^{10}\; \textrm{M}_\odot$ \citep[see figure 10 in][]{Padmanabhan2004}, a factor $\sim 40$ lower than 
 the global SHMR. 
 
 Given the difficulty in inferring the virial mass from the kinematics of the inner region alone \citep[e.g.][]{Mamon2005}  and given the significant effects it can have on the DM distribution in small-mass galaxies \citep[e.g.][]{Pontzen2012,Dutton2016}, it is surprising that  the ultra-diffuse  galaxy (UDG) \NDF\  has attracted 
a great deal of attention as a galaxy with little DM.

The stellar velocity dispersion, $\sigma_*$, in \NDF\ has  been measured  separately by \cite{Emsellem2018}  using the MUSE integral-field spectrograph at the (ESO) Very Large Telescope
 and by \cite{Danieli2019} with the Keck Cosmic Web Imager (KCWI).
The velocity dispersion measurements are sufficiently different that they could lead to 
contrasting implications regarding the mass of the galaxy. 
Assuming that \NDF\ is at a distance $D=18$ Mpc, the value $\sigma_*=8.5^{2.3}_{-3.1} \kms$  derived by \cite{Danieli2019}  yields a dynamical mass $M(<1.3R_\textrm{e}=(1.3\pm 0.8) \times 10^8 \; \textrm{M}_\odot$.
However, \cite{Emsellem2018} obtain $\sigma_*=10.8^{+3.2}_{-4} \kms$, giving  $M(<1.8R_\textrm{e})=3.7^{+2.5}_{-2.2} \times 10^8$ 
and a factor of $1.6$ smaller for $M(<R_\textrm{e})$.
By extrapolating the estimate $M(<1.8R_\textrm{e})$ of \cite{Emsellem2018} out to the virial radius using the NFW density profile, we obtain a virial mass $\Mv\approx 1.4^{+2}_{-1}\times 10^{9} \; \textrm{M}_\odot$ for a concentration parameter $c=14$ and 
$\Mv =  2.7^{+5.2}_{-2.1}\times 10^{9} \; \textrm{M}_\odot$ for $c=7$. 
For the stellar mass $\Ms \sim 1.6 \times 10^8\; \textrm{M}_\odot$   of \NDF ,  the  standard SHMR
implies $\Mv \sim 6 \times 10^{10}\; \textrm{M}_\odot$, significantly larger than $\Mv$ obtained by an extrapolation of $M(<1.8R_\textrm{e})$, even using the value  in \cite{Emsellem2018}. 
This mismatch between the SHMR and kinematical mass inference  undoubtedly bears  a close similarity to the situation of ETGs discussed above. 
{In both situations and in spite of the difference in the masses, the kinematics refers to the inner regions and in both case,  the prediction of the total mass falls short of the expectations}. 
{A few points should be emphasized in relation to the SHMR.  
The SHMR gives the peak virial  mass of the progenitor throughout its entire history. For satellite galaxies, the difference between the peak and current mass can be very large (a factor of three or so)  due to  tidal stripping by the gravitational field  of the host galaxy. 
Furthermore, the  considered stellar mass is at the lowest end where the SHMR is not actually well constrained.  Another potential caveat is that the standard SHMR may not apply to  UDGs; however  numerical simulations indicate that the progenitor's halos of low- and high-surface-brightness galaxies  actually share  main properties  \citep{2019MNRAS.485..796M}.}

The current paper offers a dynamical scenario that accommodates a high $\Mv$ for the progenitor galaxy  with the observed stellar kinematics of \NDF .
We extend the idea presented in \cite{Nusser2019} dealing  with the kinematics of  10 globular  clusters (GCs) 
 in \NDF\  \citep{vanDokkum2018}. 
{In addition to  the stripping  of the outer parts of the galaxy via  external gravitational tides, the current paper explores the dynamical consequences of 
 gas removal either via  energetic feedback from supernovae or ram pressure stripping. 
 This scenario invokes standard baryonic processes that  have been demonstrated to have strong effects on the dynamics  of low-mass halos \citep[e.g.][]{Pontzen2012,Muratov2015,Dutton2016}.}

 As in any dynamical modeling, the  distance is needed to set the spatial physical scale.
There are two distance estimates   in the literature;   \cite{Trujillo2019} derive a distance  $D\approx 13 $ Mpc,
while \cite{vanDokkumDist} and \cite{2019arXiv191007529D} , respectively,  report $D=18.7\pm 1.7 $ and $18.8^{+0.9}_{-1.1}$ Mpc. 
The nearer distance of 13 Mpc  leads to a  smaller  stellar mass ($\sim 6\times  10^7 \; \textrm{M}_\odot$) and, thus, 
 brings \NDF\ closer  to  
 the standard SHMR \citep{Trujillo2019}. 
Therefore, the main interest is in the mass models for  $D= 18$ Mpc, and here, we present results for this distance only. 
 
 We define the virial radius $\rv$  to be  the radius of the sphere  within which the mean halo density is 200 times the critical density
  $\rho_\mathrm{c}=3H_0^2/8\pi G$.
  We adopt cosmological parameters based on the recent Planck collaboration \citep{PlanckCollaboration2018}. Throughout this paper, we take the Hubble constant $H_0=67.8\kms\; \mathrm{Mpc}^{-1} $ and the baryonic and total mass 
  density parameters $\Omega_\textrm{b}= 0.049$ and $\Omega_\textrm{m}=0.311$.
  
 The outline of the remainder  of the paper is as follows. 
Our   analysis  relies on numerical simulations run under the assumption of a spherical configuration. The numerical setup and the modeled baryonic processes are described in \S\ref{sec:setup}.  The inferred prediction for the line-of-sight (LOS) velocity dispersion and comparison with the 
observed dispersion is presented in \S\ref{sec:results}. 
We conclude with a summary and a discussion in \S\ref{sec:con}.

\section{The setup} 
\label{sec:setup}

We start with a progenitor (satellite) halo of virial mass $M_\textrm{h}$ of a few times $10^{10}\; \textrm{M}_\odot$ at the outskirts  of  a larger parent galaxy at redshift $z\sim  2-3$. 
Since the cooling time in this halo-mass range is short compared to the dynamical time \citep[e.g.][]{WF91},  we 
assume that a significant 
gas fraction has  already cooled and settled well inside the virial radius.  
We also assume that at this stage,  only a small 
 fraction \ltsim 0.2 of the final stellar mass has formed
  \citep{Behroozi2013a,Kimmel2019}.  {We require that most star-formation and its associated feedback occur 
 at later stages of the evolution.}
 
As the (gas-rich) satellite orbits  into the parent halo, it  begins to 
 lose matter from its outer parts by the tidal gravitational forces of the parent halo. It also continues to 
form stars accompanied by supernova (SN) explosions, perhaps at a boosted rate  due to tidal interactions with the host halo \citep{Martig2008,RBK14}. 
Because  the gravitational potential depth in the satellite halos we consider corresponds to circular velocities $\vc\ltsim 60\kms$, SN feedback  is 
 sufficient for the  removal of significant amounts of gas away from the gravitational grip of the halo \citep{Larson74,DS86,Munshi2013}. 
{ For the relevant mass range,  simulations of galaxy formation  \citep{Muratov2015} indicate that the amount of gas ejected could be more than an order of magnitude larger than the mass in forming stars.  
 In addition to star-formation feedback,  another mechanism for gas removal is  ram pressure exerted by  the diffuse gas in the parent halo, but it is hard to estimate this effect given the little we know about the parent galaxy, especially at earlier times. 
 Nonetheless, the UDG  \NDF\ is extremely gas poor  \citep{Chowdhury2019,Sardone2019}
 and it is natural to assume that all of the gas that has not turned into stars has been ejected. The actual causes of gas removal are not important to our modeling, since we are only concerned with its dynamical effects. 
  }

{Continuous star-formation activity  could lead to slow removal of gas on time scales longer than the dynamical time of the system, while a starburst  could cause a fast removal of the gas. }
{Gas removal by ram pressure could be either or fast  or slow depending on the orbital speed of the satellite and the 
gas density on the parent halo. }
 In  slow gas removal, the halo  DM particles remain in a quasi-steady state. In contrast,  in a  fast ejection process, the DM particles momentarily maintain their velocities, while gaining  (positive) potential energy. 
After a few dynamical times,  the system  relaxes to a new  equilibrium, which, in general, is different from the 
steady state reached at the end of  a slow gas-removal process \citep[e.g.][]{Pontzen2012,Dutton2016}. 

 {At at a distance $D=18$ Mpc,  \NDF\ is  at a projected distance of $\approx  75 $ kpc from the 
the large elliptical NGC 1052 and with a relative LOS speed of $293 \; \kms$ \citep{vanDokkum2018}.  We emphasize here that based on the measured LOS velocity dispersion in NGC 1052 ($\sim 110 \; \kms$),  the relative speed of 
\NDF\ is close to the escape velocity from NGC 1052 \cite[see][]{Nusser2019} and the two galaxies, \NDF\   are  likely to be just skimming each 
other. Still, the gravitational tides of NGC 1052 could certainly be strong enough to cause significant mass  
mass loss in \NDF .  
A  larger mass for  NGC 1052 is  obtained from   the SHMR relation. 
The   NGC 1052 stellar mass  of  $\sim 10^{11}M_\odot$ \citep{2017MNRAS.464.4611F}, translates 
to a 
 halo  mass of $\sim 5\times 10^{12}M_\odot $ \citep{Wasserman2018} according to the SHMR.
 In any case, there is a large uncertainty in the estimation of the tidal radius of \NDF\ in the vicinity of NGC 1052 \citep{Ogiya2018, Wasserman2018}.  Nevertheless,  a tidal radius of  $\sim10$ kpc is consistent with  the observed  spatial extent of the stellar component of \NDF\ and the distribution of the projected distances of its star clusters. 
 }
 
\subsection{The Numerical Scheme} 
\label{sec:nscheme}
Under the assumption of a spherically symmetric configuration, we simulate the dynamical effects of cooling, stripping, and gas ejection.
Only the collisionless  particles are ``live" and move self-consistently under the action of  their own gravity as well as that of the stellar and cool-gas components. 
The gravitational force field of these baryonic  components is 
computed assuming they follow  the 
density profile of the observed stellar component
but with a mass that  varies with  time according to the cooling and galactic wind recipes described below. 
Thus, the distinction between the stars and the cool gas is unimportant dynamically and, at times, we shall refer to them as accreted baryons.

Initially, the collisionless particles represent the DM and the hot gas, where the latter is assumed to follow the density distribution of the DM halo.
The collisionless particles are treated as spherical shells moving under the gravitational force field of the monopole term of the mass distribution computed relative to the halo center (as defined in the initial conditions). 
The self-gravity of the collisionless particles is derived  following  \cite{White1983}, with a force Plummer softening of length of 10 pc.
{The time integration is performed using the leapfrog scheme with a variable time step chosen to be 0.05 the dynamical time scale at the 
center of the halo. }

The numerical scheme aims at modeling  the dynamical consequences of the following  processes: 
\begin{enumerate}[label=\Roman*.]
\item \noindent  \textit{Gas cooling:}  the mass of the cool gas, $\Mg$, is increased linearly from zero to $\Mg=\fg \Mv$ in the time period from $t_\textrm{C1} =1 $  to $t_\textrm{C2} =2 $ Gyr. The parameter  $\fg=\Omega_\textrm{b}/\Omega_\textrm{m}$ is the gas  fraction of the total halo mass. 
The mass of each collisionless particle is accordingly reduced by a factor $1-\Mg(t)/\Mv$.
The distribution of the cool mass, $\Mg(t)$, is set  according to the density profile of  the observed stellar profile described below.
\smallskip
\item \noindent \textit{Trimming/Stripping:} 
\begin{enumerate}[label=\alph*.]

\item \noindent \textit{Linear:} Collisionless particles lying at time $t$ beyond a trimming radius $\rtr(t)$ are excised from the simulations, where 
$\rtr(t)$ varies linearly from $\rtr=\rv$ at $t_\textrm{tr1}=t_\textrm{C2}=2$ Gyr to $\rtr=10$ kpc at $t=t_\textrm{tr2}=4$ Gyr. 

\item \noindent \textit{Dynamical tidal stripping:} the actual gravitational field of the parent halo is included in the simulation.
The satellite is assumed to move on a circular orbit and the equations of motion are solved in the non-inertial frame rotating with the satellite, taking into account the Coriolis and centrifugal forces in addition to the gravity of the parent halo. 
As before, the gravitational force of the satellite  is computed assuming spherical symmetry with respect to its centre. The approximation of spherical symmetry is unrealistic for the already stripped particles forming  galactic streams. However, the dynamics of those  particles is irrelevant to us. 

\end{enumerate}

\smallskip
\item \noindent \textit{Galactic winds:} 
\begin{enumerate}[label=\alph*.]
\item \noindent \textit{Fast:} a mass $\Mg=\fg \Mv-\Ms$, where $\Ms$ is the observed mass of the stellar component,  is suddenly removed from the accreted baryons at  $t=t_\textrm{w1} =t_\textrm{tr2}=4$ Gyr.
The remaining mass  $\Ms $ is fixed thereafter and represents the observed  stellar component. 
\item \noindent \textit{Slow:}  a mass $\Mg=\fg \Mv-\Ms$ is removed linearly with time between 
$t=t_\textrm{w1} $ and $t_\textrm{w2}=8$ Gyr. 
\end{enumerate} 
\end{enumerate}

For reference, the period of a circular orbit of radius (in kpc) $r_\textrm{kpc}$ is 
$
 t_\textrm{c} =2\pi ({r^3}/{GM})^{1/2}= 0.3({r_\textrm{kpc}^3}/{M_8})^{1/2}\textrm{Gyr}
 $ where $  M_8 $ is the mass (in $10^8\; \textrm{M}_\odot$)  within $r_\textrm{kpc}$. 
For $M\sim 10^8\; \textrm{M}_\odot $ at $r\sim 2$ kpc giving $t_\textrm{c}\sim 0.85 $ Gyr for the orbital period of a
circular orbit at $r\sim 2$ kpc inside  \NDF . 
The circular orbital period in the parent galaxy is 
$
t^\textrm{DF2}_\textrm{c}\sim  2.5 ({R_{100}}/{V_{250} }) \;  \textrm{Gyr} ,
$
where  $R_{100}$ is the orbital radius in 100 kpc  and  $V_{250} $ is the velocity in $250 \;  \kms$.

The approach outlined above is obviously quite limited in comparison to 3D  simulations of galaxy formation, but  it has the advantage that the dynamical consequences of the baryonic processes   can easily be assessed and discerned. 
Full 3D simulations including gas process, star formation, and feedback are more realistic, but they suffer from uncertain complex sub-grid modeling and are substantially harder to analyze.

\subsection{The stellar density profile}

\begin{figure*}
  \includegraphics[width=0.96\textwidth]{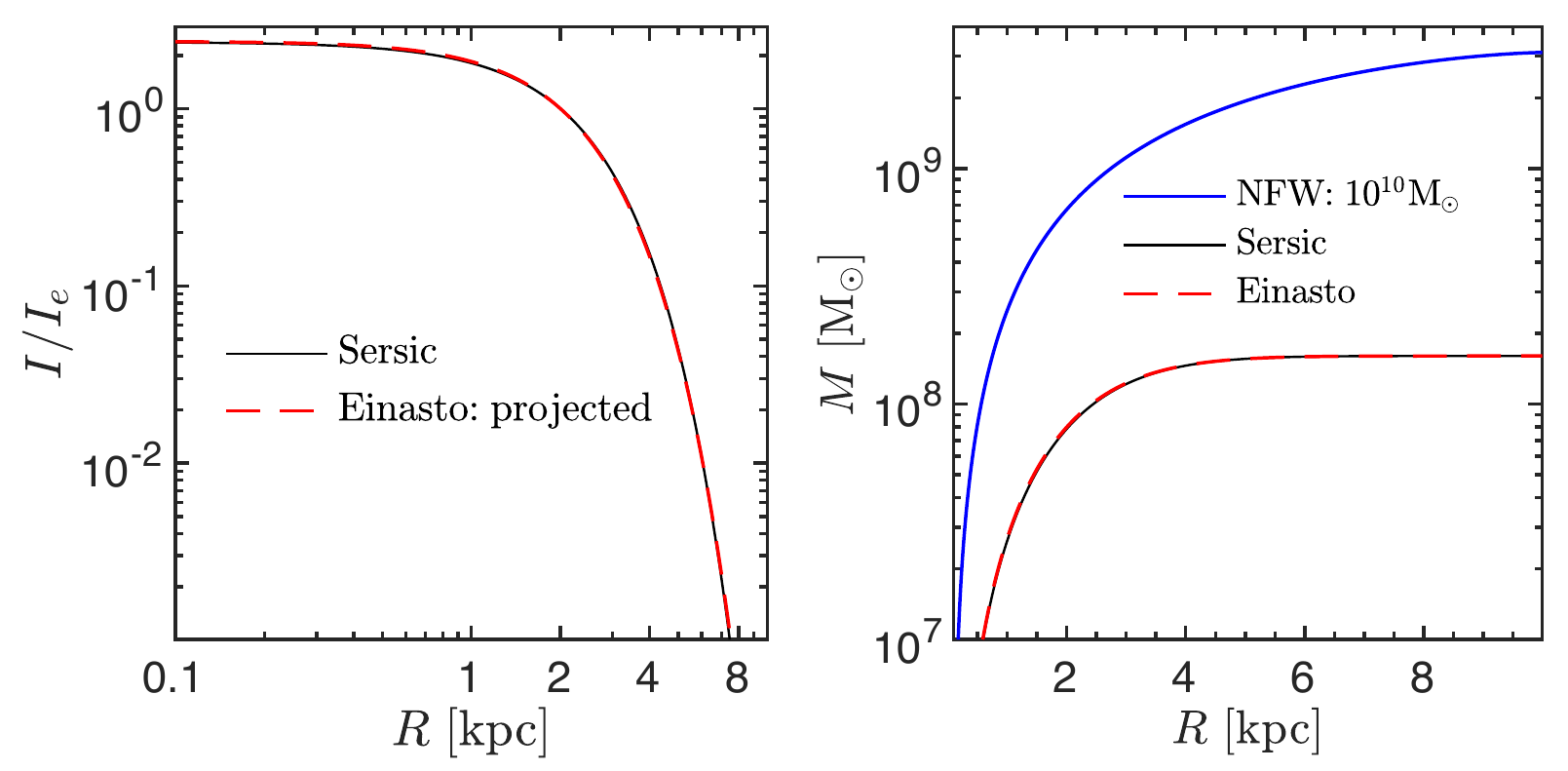} 
  \vskip 0.0in
 \caption{ {Left panel:} the  stellar surface density versus projected distance $R$. The S\'ersic fit to the observed \NDF\ 
 stellar distribution is shown as the solid black line, while the overlapping dashed red line is the projected 
 3D Einasto profile   truncated at 10 kpc. 
 {Right panel:} stellar mass  within 2D cylinders versus $R$. The solid blue line correspond to an NFW halo with $\Mv= 10^{10}\; \textrm{M}_\odot$ 
 and $c=9$ truncated at 10 kpc. }
\label{fig:Serc}
\end{figure*}

We describe  the 3D density distribution of the observed stellar component in terms of 
 an Einasto profile
 \begin{equation}
 \label{eq:rhoobs}
 \rhoobs=\rho_0 \exp\left[-\left(\frac{r}{h}\right)^{1/n}\right] \; ,
 \end{equation}
where $n=0.649$, $h=r_{-2}/(2n)^n$,   $r_{-2}=2.267$ kpc and $\rho_0$ and $\rho_0$ is tuned to yield the total mass. 
This profile  yields a good fit to the 2D S\'ersic  profile representing the observed surface brightness  with 
 $R_\textrm{e}=2 $ kpc (for $D=18$ Mpc) and a S\'ersic  index $n=0.6$ \citep{vanDokkum2018}. The stellar mass is
 is normalized to $1.6\times 10^8 \; \textrm{M}_\odot$.
The left panel of Fig.~\ref{fig:Serc} plots the   S\'ersic  profile as a function of the projected distance, $R$, together with the surface density obtained by integrating the Einasto profile along the line of sight out to a maximum 3D radius of 10 kpc. The agreement between the two profiles is excellent. To the right we show the mass enclosed in cylinders of radius $R$ for these two profiles. In addition, we show here the 2D  mass obtained with an NFW profile of virial mass 
$\Mv=10^{10} \; \textrm{M}_\odot$ and $c=9$ pruned at a 10 kpc 3D distance. 

\subsection{Derivation of the stellar velocity dispersion}
\label{sec:Jeans}
Our goal is  predicting  the line of sight velocity dispersion as a function of the projected distances. 
The simulations provide  the total gravitational force field (per unit mass), $g(r)$,  resulting from the DM and baryonic components. 
Since the  stellar component is represented in terms of a fixed density profile,  
we resort to the Jeans equation in order to derive the velocity dispersion of the stars.

Let $\sigma_r^2(r)=\langle v_r^2 \rangle$ be the stellar velocity dispersion in the radial direction
at distance $r$ from the center of the galaxy. The velocity anisotropy ellipsoid is described 
by the parameter $\beta(r)=1-\sigma_t^2/2\sigma_r^2$ where $\sigma_t^2 =\sigma_\phi^2+\sigma_\theta^2$ is the 
velocity dispersion in the direction  tangential to $r$.
 The Jeans equation for a steady-state distribution of stars is
 \begin{equation}
 \frac{\dd n \sigma_r^2}{\dd r}+\frac{2\beta} {r} n \sigma_r^2=-n g
 \end{equation}
 where $n=n(r)$ is the known number density of stars at $r$.
 Assuming $\beta$ is independent of $r$, the solution to this equation is 
 \begin{equation}
 \sigma_r^2=-\frac{1}{n(r) \; r^{2\beta}}\int_r^\infty n(r')\;  r'^{2\beta}g(r') \dd r'\; .
 \end{equation} 
  
 There is no restriction on $\beta$ in the expressions above, but there is no guarantee that a particular $\beta$ actually corresponds to any physical system described by a steady-state phase space distribution function (DF). 
 Indeed,  numerical integration of  Eddington's formula did not yield a nonnegative  isotropic  DF (i.e. $\beta=0$); 
   \citep[see the discussion in \S4.3.1 in][]{Binney2008} that describes
   the observed stellar distribution (modeled as an Einasto profile)  in equilibrium under its self-gravity and  the gravity of an NFW halo.  For  $\beta<-0.01$, models of the form $f(E)=L^{-2\beta}f_1(E)$ are  numerically found to be consistent with the desired configuration. We thus restrict ourselves to negative $\beta$\footnote{Including only 
    the self-gravity of the stellar component allows for 
 an    isotropic and nonnegative DF. But  for $\beta>0$, 
the models $f(E)=L^{-2\beta}f_1(E)$ lead to  negative $f$. 
 }.
 
We choose the line of sight to lie in the $z$-direction.
The  velocity dispersion, $\sigma_z^2=\langle v_z^2\rangle$, at  a point $R $ and $z$
is 
\begin{equation}
\sigma_z^2=\sigma_r^2\left(1-\beta \sin^2\theta\right)\; ,
\end{equation}
where $\sin\theta=R/\sqrt{R^2+z^2}$. The observed parallel velocity dispersion at a projected distance  $R$ 
is 
\begin{equation}
\sigma^2_u=\frac{1}{\int_\infty^\infty \dd z \; n(z,R)}\int_\infty^\infty \dd z\;  n(z,R) \sigma_z^2 \;.
\end{equation}
Comparing to the observed velocity dispersion, we show (unless stated otherwise) results after  averaging  $\sigma_u^2$
over the stellar  surface density within $R$. 

\section{Results}
\label{sec:results}

The initial mass in stars and cool gas  in the progenitor satellite are set to zero,
and the initial positions and velocities of the collisionless particles are sampled  from an ergodic (i.e. isotropic) phase-space DF.
The DF is numerically derived from the NFW density profile  by integrating the Eddington relation. This DF is a monotonic function of energy,  and thus, the system is stable
\citep{Binney2008}. The simulations are run for a period of $9.7$ Gyr. In order to reduce effects related to the finite number of particles, the simulations  are first run for 
 1 Gyr, before  the sequence of baryonic processes is   implemented ({I}--{III} above).
 The energy conservation of the code for  $10^6$  self-gravitating collisionless particles alone is better than 1 part in $10^4$.
 
{
We will explore the consequences of the processes discussed in  \S\ref{sec:nscheme} for 
 halos with NFW density profiles corresponding to virial masses,  $\Mv=10^{10}\; \textrm{M}_\odot$ and  $5\times 10^{10}\; \textrm{M}_\odot$ at $ z=0$ and  $\Mv=2\cdot 10^{10}\; \textrm{M}_\odot$
 at $z=2$. }
   The virial radius and circular velocity corresponding to  $\Mv=10^{10}\; \textrm{M}_\odot$  are $46$ kpc and $31\kms$, respectively. For this halo, 
 we consider two values of the concentration parameter, $c=9$ and $13$.
   Using the \cite{Ludlow2016a} recipe for calculating the halo concentration versus halo mass (hereafter, the $c\! -\! M$ relation), 
the mean concentration for halos of this mass is $c=12.4$. 
The value $c=9$  is  within the $1\sigma$ scatter of 
the expected variations of $c$ in relaxed halos  \citep[see Fig.~5 in][]{Hellwing2016b} and is motivated by the analysis of \cite{Nusser2019}, which found that  the GC kinematics prefers smaller concentrations. 
For  the larger  halo $\Mv=5\times 10^{10} \; \textrm{M}_\odot$ we consider $c=8$  and $c=11$, where 
the   $c\! -\! M$ relation gives $\bar c=11.1$ and $c=8$ is  within the  $1\sigma$ scatter. 
 The virial radius and the circular velocity of this halo are $78$ kpc and $53 \kms$, respectively.
{
The  $z=2$ halo with $\Mv=2\cdot 10^{10}\; \textrm{M}_\odot$ actually matches the median mass of the progenitor of the  $z=0$ 
$\Mv=5\cdot 10^{10}\; \textrm{M}_\odot$ halo \citep{Correa2015}. For the Planck cosmology, the critical density,   $\rho_\mathrm{c}=3H_0^2/8\pi G$ is a factor of nine  larger at $z=2$, yielding a 
  virial radius of $28$ kpc for  $\Mv=2\cdot 10^{10}\; \textrm{M}_\odot$  at $z=2$.  Assuming the evolution of the halo is close to stable clustering, the density profile of the 
$z=2$ halo should agree with   $ z=0$ halo profile  at $r<28$ kpc. Indeed, 
the mass within a radius of 28 kpc in the large $z=0$ halo is 
 $2.4\cdot 10^{10}\; \textrm{M}_\odot$,  close to the mass of the $z=2$ halo. According to the recipe of \cite{Ludlow2016a}, the median  concentration of the $z=2$ halo is $c=6$. We therefore  find
$c=3$ to be  the concentration of this halo at which we find consistency  with the expected scatter in the 
 $c-M$ relation. }
The number of collisionless particles in the simulations with the high-mass halo is $10^7$ and $10^6$ for the two lower-mass halos.

\subsection{Small halo}
\begin{figure*} 
  \includegraphics[width=.96\textwidth]{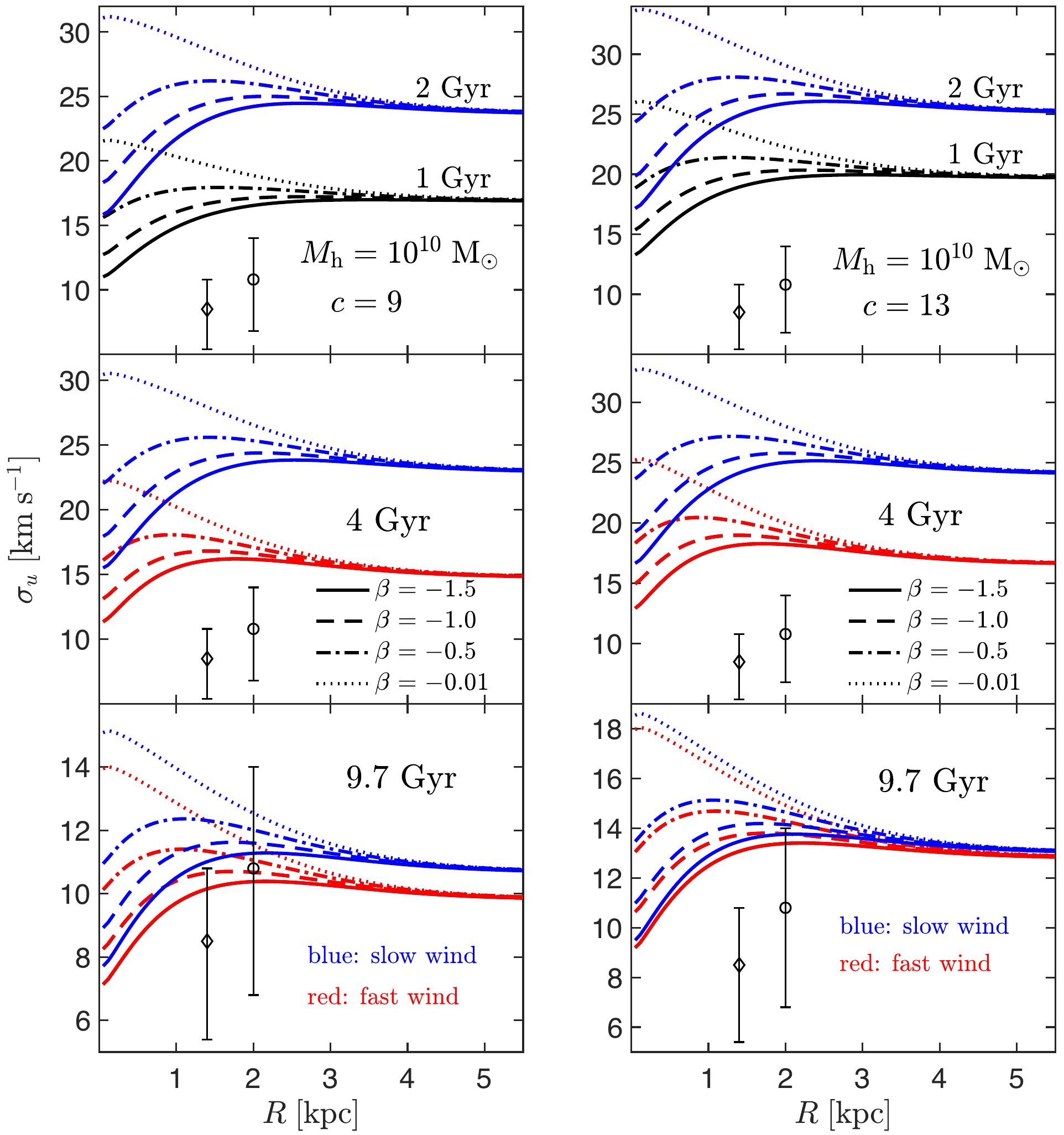} 
  \vskip 0.0in
 \caption{ The LOS velocity dispersion versus projected distance obtained from the simulations with virial halo mass $M_\mathrm{h}= 10^{10} \; \textrm{M}_\odot$ and $c=9$ (column to the left) and $c=13$ (to the right). 
 Each group of four curves represent  four values of the velocity anisotropy parameter, $\beta$, as indicated in the middle row. 
 Red and blue correspond to fast and slow winds, respectively. At $t=2$ Gyr, the red and blue curves overlap.
 The black curves in the  top panel shows results at $1 $ Gyr just before the gas is collapsed. Shown are results from simulations without tidal stripping/trimming. The diamond and open circle  are the \citep{Danieli2019} and \citep{Emsellem2018} measurements, respectively. The value $\fg=0.155$ is assumed.}
\label{fig:twoC}
\end{figure*}

We begin with   $\Mv=10^{10}\; \textrm{M}_\odot$. 
Using the Jeans equation, we calculate $\sigma_u$ from the simulation output as described in
 \S\ref{sec:Jeans}. 
Fig.~\ref{fig:twoC} plots curves of $\sigma_u$  versus the projected distance $R$ 
for  $c=9$ (left column) and $c=13$ (right). Each group of curves corresponds to four values of $\beta$, as indicated in the middle row panels. 
The group of black curves in the top row  panels are computed from the simulations at  $t=1$ Gyr (black curves), 
just before gas cooling is switched on, as described in \S\ref{sec:nscheme}. At this stage, neither the stellar nor the cool-gas component  have formed, but we still compute  $\sigma_u$ 
for a population of ``massless" tracers distributed according the observed density profile $\rhoobs$ given in Eq.~\ref{eq:rhoobs}. 
The diamond and open circle  with attached error bars represent the observed velocity  dispersions from
\cite{Danieli2019} and \cite{Emsellem2018}, respectively.  These points are plotted at different projected distances as given in those papers. 
This figure refers to results without any trimming/stripping.

At 2 Gyr (blue curves, top panels), the cooling phase ended with  a cool-gas mass $\Mg=(\Omega_\textrm{b}/\Omega_\textrm{m})\Mv=1.55\times 10^9 \; \textrm{M}_\odot $
assumed to follow the form $\rhoobs$. There is a significant enhancement of $\sigma_u$ as a result of the transfer of mass from the collisionless particles to
 the more centrally concentrated   baryonic  component.  During the  period 2-4 Gyr,  
 the baryonic component remains the same and since no trimming is invoked,  the distribution of the collisionless particles essentially remains the same.

The red curves in  middle panel show  $\sigma_u$  just after an event of fast wind ejecting in a single burst a mass  $\Mg-\Ms= 1.39\times 10^{9}\; \textrm{M}_\odot$ 
corresponding to all available cool gas after  leaving behind the stellar component with $\Ms=1.6\times 10^8\; \textrm{M}_\odot$. In the same panels, 
the blue curves correspond to a slow wind  lasting  until $t_\textrm{w2}=8$ Gyr (see III.b in \S\ref{sec:nscheme}). 
Since the slow wind begins at $t_\textrm{w2}=4$ Gyr, the blue curves in the middle and top panels are almost identical, with small differences 
entirely due to fluctuations in the distribution of the collisionless particles. 

By $t=9.7$ Gyr, the system reaches a steady steady with a galaxy made of DM collisionless particles in addition to   stellar component of mass $M_*$. The results for $c=9$ in the bottom panel on the left indicate that steady state is sensitive to the gas removal mode. The fast wind is more efficient at
bringing the velocity dispersion to a comfortable agreement with the observations. 
As evident in the bottom panel in the column to the right, for the larger concentration, $c=13$, the final $\sigma_u$ is consistent with the measured dispersion at about the $2\sigma$ level for $\beta=-1$ and $\beta=-1.5$. For this more concentrated halo, the two wind modes yield roughly similar results. 

\begin{figure*} 
  \includegraphics[width=0.96\textwidth]{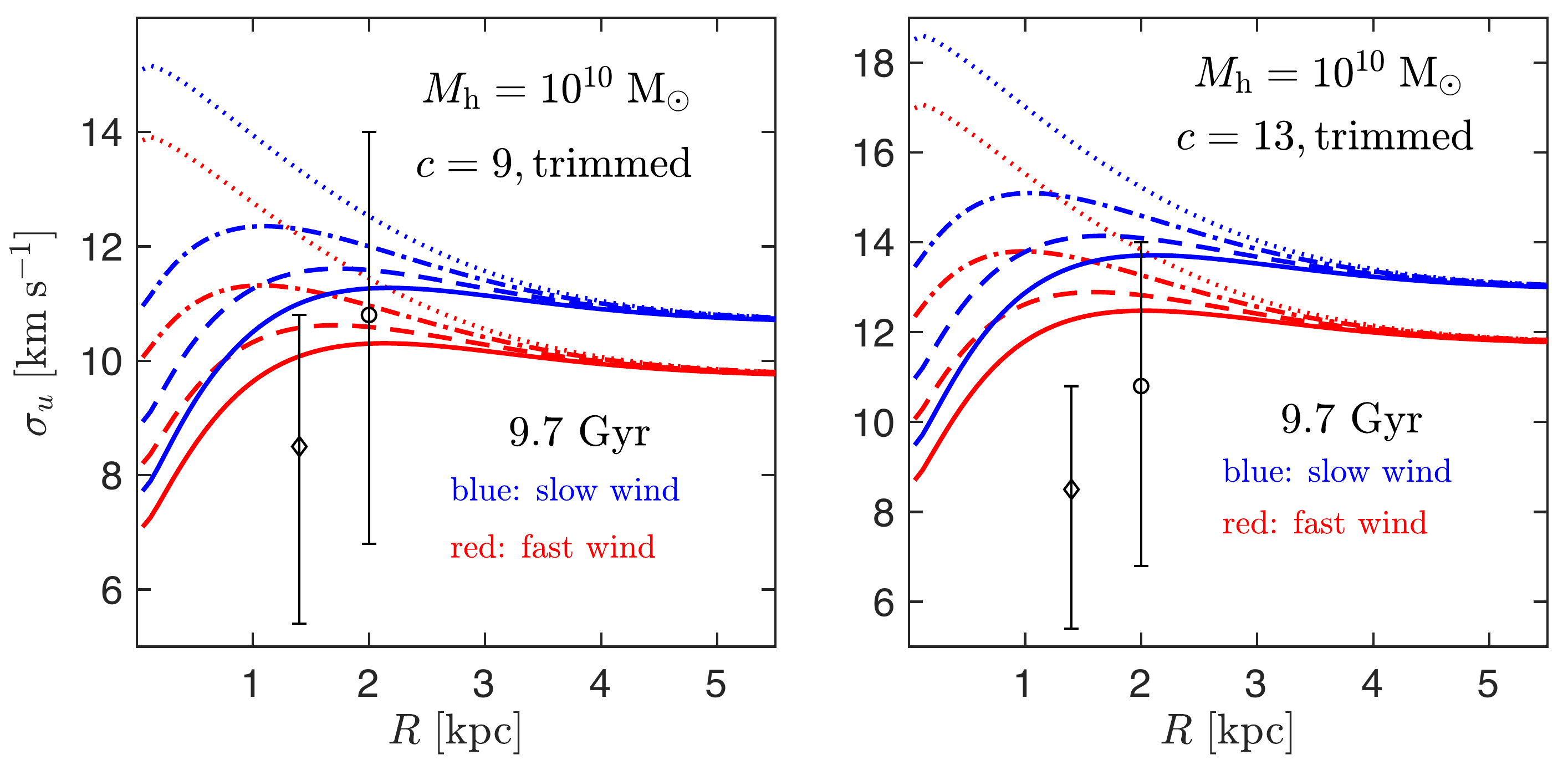} 
  \vskip 0.0in
 \caption{The same as the bottom panels in the previous figure but for the simulations with trimming. }
\label{fig:twoCtrm}
\end{figure*}

\begin{figure}
 \includegraphics[width=0.48\textwidth]{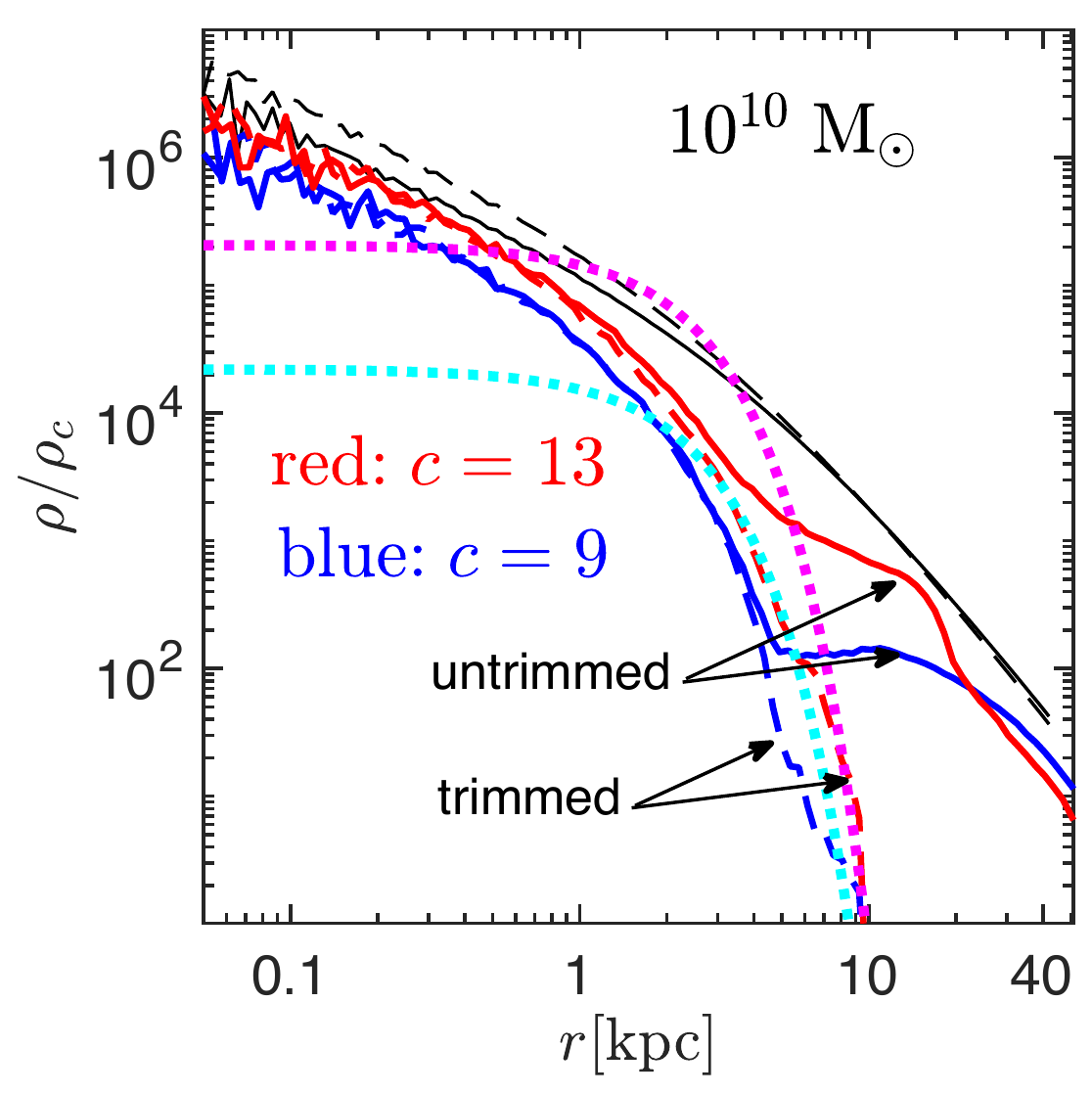} 
  \vskip 0.0in
 \caption{Density profiles (in units of the critical cosmic density) obtained from the simulations with fast wind for $\Mv=10^{10}\; \textrm{M}_\odot$.   Dashed and solid thin black are  the DM density profiles, respectively, for $c=13$ and $c=9$, at $t= 1$ Gyr just before turning on gas cooling. 
 The DM profiles at $t=9.7$ Gyr  without trimming are
    thick solid, where the red is for $c=13$ and blue for $c=9$. 
 The thick dashed curves are the same as the solid, but for simulations with linear trimming.  The dotted blue curve is the stellar profile at the final time, while the dotted red is the baryonic density profile at $t=2$ Gyr, i.e. the maximum value reached by $\Mg$. 
 }
\label{fig:rhoplt}
\end{figure}

\begin{figure}
 \includegraphics[width=0.48\textwidth]{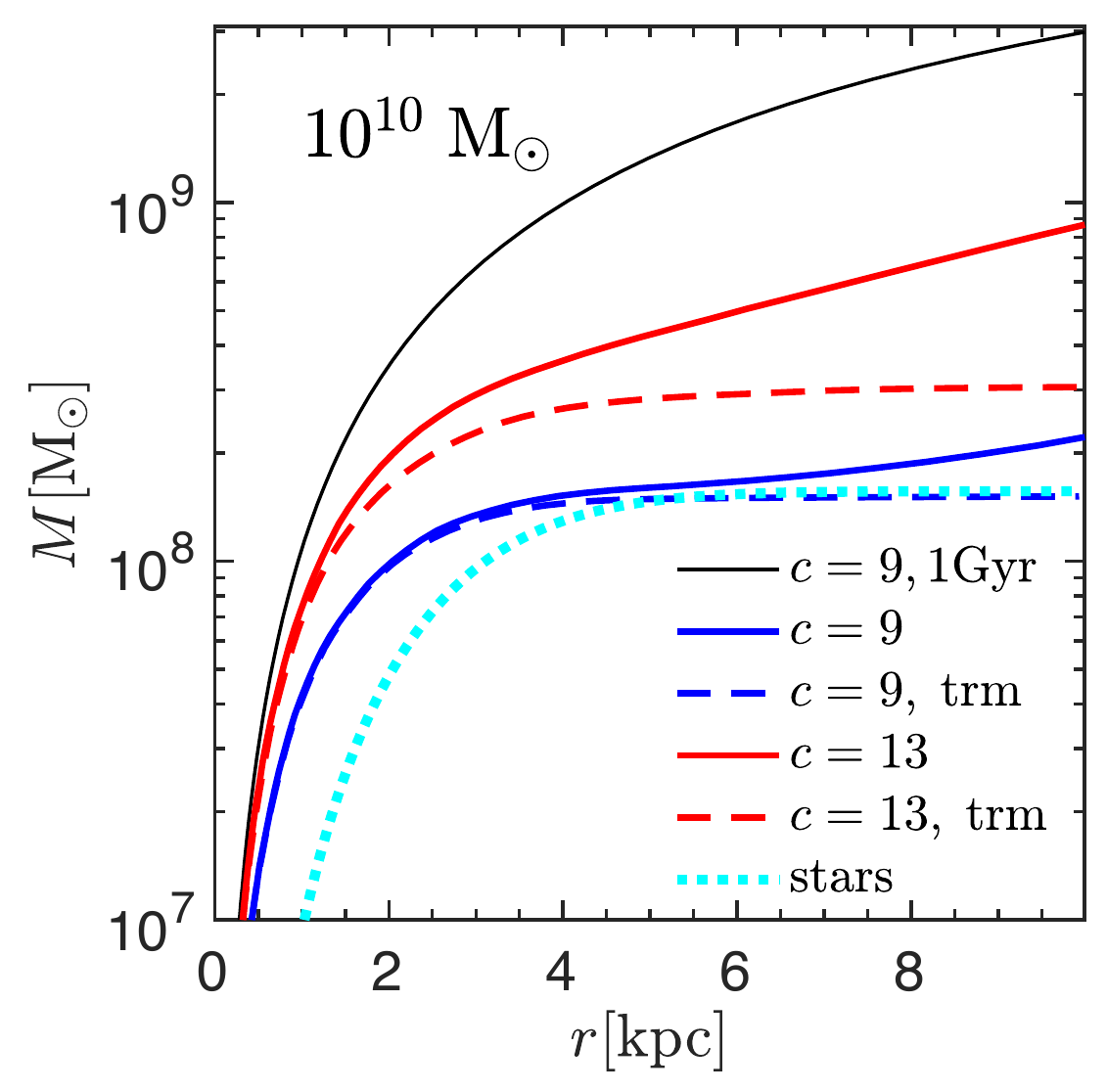} 
  \vskip 0.0in
 \caption{ Mass enclosed in a 3D radius $r$ for several   of the density  profiles from the previous figure. }
\label{fig:Mr}
\end{figure}

Fig.~\ref{fig:twoCtrm} explores the effect of linear trimming (II.a in \S\ref{sec:nscheme}). 
For brevity, the plot shows the results in the steady state limit at $t=9.7$ Gyr.   Trimming has very little effect on $\sigma_u$ for the simulations with $c=9$, as readily seen by comparing this figure and the bottom panel in the previous figure. For  $c=13$, the effect of trimming is more pronounced for the fast wind 
mode, while almost negligible for slow wind.  To understand this behavior, we inspect in Fig.~\ref{fig:rhoplt}  the actual density profile for various cases. 
The solid  and dashed  curves in blue ($c=9$)  correspond to densities of DM particles at the final time for untrimmed and trimmed simulation runs, respectively.
The curves almost overlap out to $r=4$ kpc and are below the observed stellar density (dotted cyan) in the range 2-5 kpc.
This explains why  trimming almost has no effect on $\sigma_u$ for $c=9$. Further, the gravitational effects of gas ejection are clearly visible in the DM profile for the untrimmed case at $r$ larger than a few kpc. 
However, for $c=13$,  the red curves deviate at $r\gtsim 1 $ kpc and both are above the observed stellar density.   
The same conclusions can be reached from Fig.~\ref{fig:Mr} showing the mass within  a radius $r$. 
The differences in  $M(r)$ between the trimmed and the untrimmed simulations is clearly more pronounced for $c=13$ (red) than $c=9$ (blue).

Even for the larger concentration, provided $\beta$ is sufficiently low,  the  model curves are consistent with the data at less that than the $2\sigma $ and $1\sigma$ for the \cite{Danieli2019} and \cite{Emsellem2018} points, respectively.

\subsection{Large halo}
We now turn to  the larger halo with $\Mv=5\times 10^{10}\; \textrm{M}_\odot$. In this case,  the  gas mass that can collapse to the inner regions is five times larger  than in the $10^{10}\; \textrm{M}_\odot $ halo. 
Thus, despite the deeper potential well ($\vc=52\kms$ compared to $31 \kms$), the fraction of gas to original DM mass within 10 kpc  is actually larger than in the smaller halo, 1.02 versus 0.514 for $\Mv=10^{10}\; \textrm{M}_\odot$ with $c=9$.

\begin{figure}
 \includegraphics[width=0.48\textwidth]{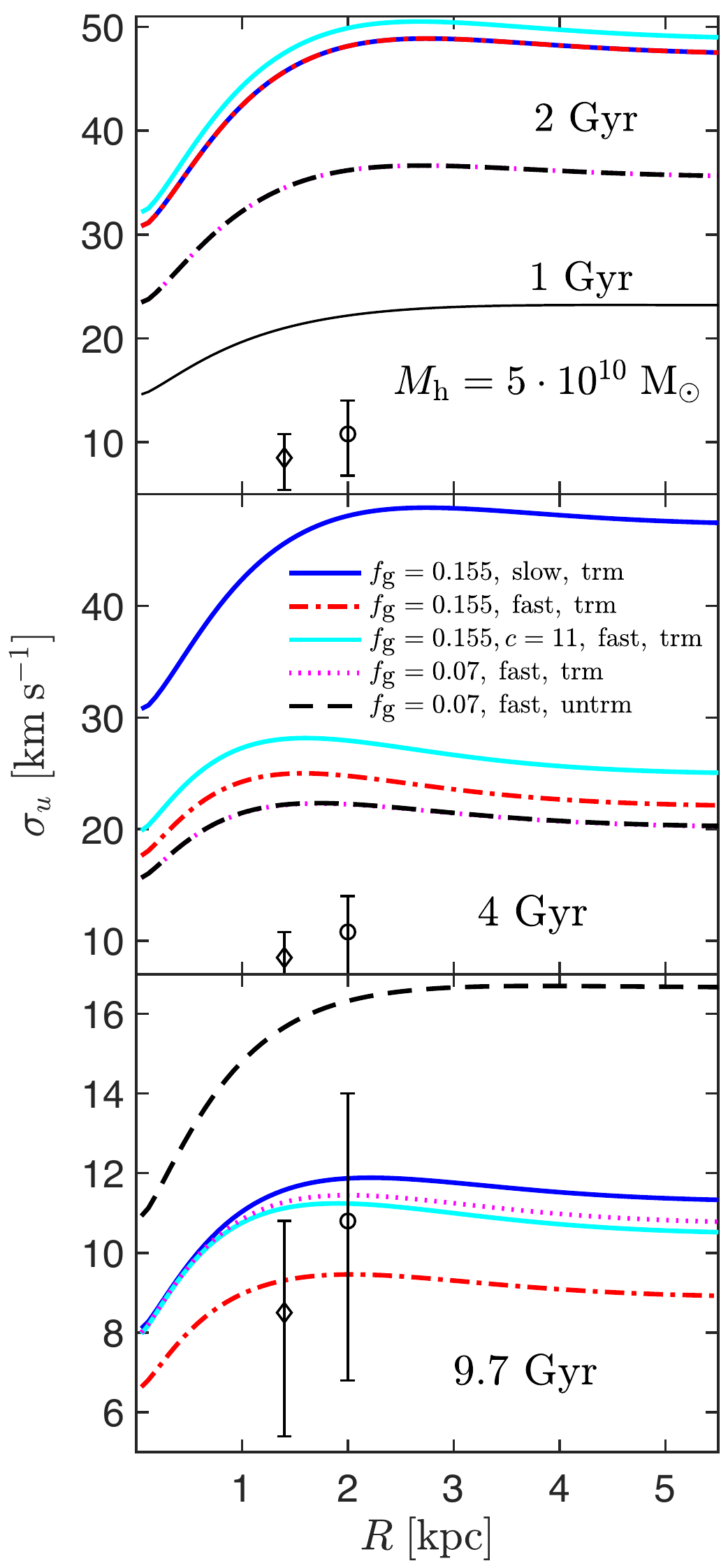}
  \vskip 0.0in
 \caption{ The LOS velocity dispersion  from the simulations with $M_\mathrm{h}=5\times 10^{10} \; \textrm{M}_\odot$ and $c=8$, with the exception of the cyan solid line, which is 
 for $c=11$. 
Only results  for $\beta=-1.5$ are plotted.  In  the top panel, 
 curves with the same $\fg$ overlap. In the middle panel, just after fast gas ejection at 4 Gyr, despite the trimming, the curves with $\fg=0.07$ are almost identical, but diverge by 9.7 Gyr as seen in the bottom panel.  }
\label{fig:bigM}
\end{figure}

\begin{figure}
 \includegraphics[width=0.48\textwidth]{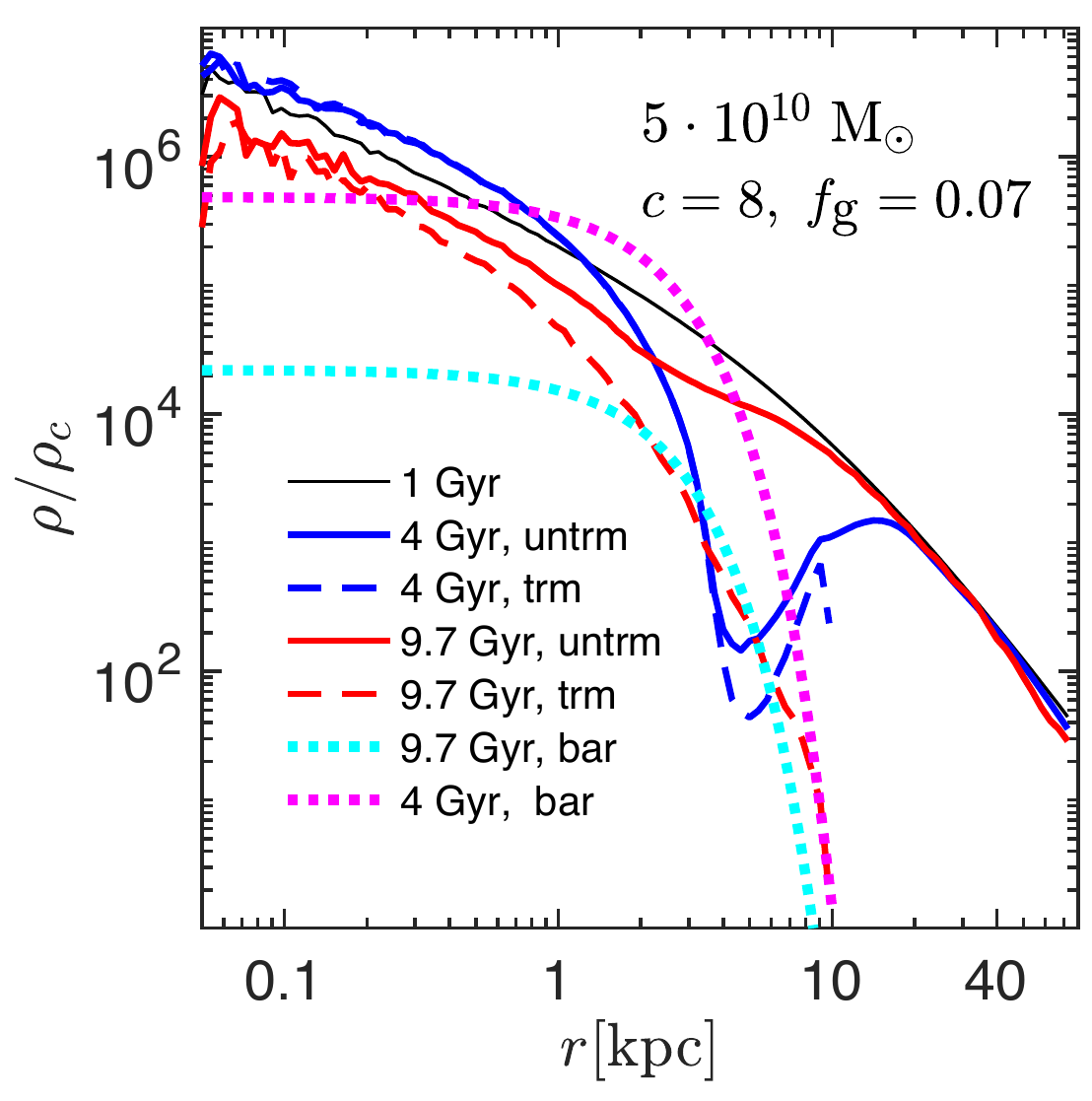}
  \vskip 0.0in
 \caption{Density profiles  from the simulations with $\Mv=5\times 10^{10}\; \textrm{M}_\odot$ with $c=8$ for fast winds.  The dotted blue curve is the stellar profile at the final time, while the dotted red is the baryonic density profile at $t=2$ Gyr, i.e. the maximum value reached by $\Mg$. 
 }
\label{fig:rhplt78}
\end{figure}

The results are shown in Fig.~\ref{fig:bigM}  for  the   cool-gas mass  fraction, $\fg=\Mg/\Mv=\Omega_\textrm{b}/\Omega_\textrm{m}=0.155$ and also for nearly half  this value, $\fg=0.07$. To avoid cluttering of the figure, we show results obtained with $\beta=-1.5$ only and simply point out that at $R\approx 0$, the velocity dispersion for $\beta=-0.01 $ is nearly a factor of two larger than 
$\beta=-1.5$. The cyan solid line corresponds to $c=11$ falling on  the $c\!-\!M$ relation \citep{Ludlow2016a}, while all other lines are for $c=8$. 
The bottom panel demonstrates that the ejection of the relatively large amount of gas for $\fg=0.155$ brings down  $\sigma_u$ (dashed-dotted) to a level completely consistent with the observations for $c=8$. 
The agreement with the observed $\sigma_u$ is even better  than the corresponding 
$\beta =-1.5$ curve  plotted  the bottom panel in Fig.~\ref{fig:twoC}  for  the lower-mass $\Mv=10^{10}\; \textrm{M}_\odot$. 
Like in the  the lower mass case, the concentration parameter plays an important role. Comparing between the cyan solid and the red dashed-dotted lines, the velocity dispersion for $c=11$ is nearly 50\% higher than $c=8$ at $t=9.7$ Gyr.
The following conclusions are also valid for $c=11$, but for clarity of the figure, we only plot results for $c=8$.
For $\fg=0.155$, the  trimmed and  untrimmed simulations  yield very similar $\sigma_u$. Therefore, only the trimmed 
results are presented in Fig.~\ref{fig:bigM}. 
In contrast, for $\fg=0.07$, trimming has a significant effect on  the final $\sigma_u$ as revealed by  the comparison between 
 the dotted (trimmed) and dashed (untrimmed) 
curves in the bottom panel. It is intriguing, however, that the two curves are almost identical at the earlier time, $t=4$  Gyr, as shown in the middle panel (small  differences 
can be seen by  inspecting  the actual numerical value). 
To understand this behavior, we plot in  Fig.~\ref{fig:rhplt78} the density profiles obtained with $\fg=0.07$. At 4 Gyr, the  trimmed (blue dashed) and untrimmed (red solid)
almost completely overlap out to $r=3 $ kpc. The red
dotted curve shows the 
 baryons (cool gas + stars) density just before ejection by fast wind just before 4 Gyr. Thus,  between $1$ kpc and $8$ kpc   the baryons are actually dominant
 until gas removal.

For $c=8$, even the smaller $\fg$ (dotted magenta)  yields a reasonable agreement  with the observations at less than a $2\sigma$ deviation from the 
\cite{Danieli2019} measurement (diamond symbol). Note that the 
at $2$  Gyr (top panel), the gas contraction  with the larger $\fg$  boosts  $\sigma_u$ to higher values than  low $\fg$.
At $9.7$ Gyr (bottom panel),  the situation is reversed so that  the large $\fg$ having lower $\sigma_u$. This is due to the more substantial reduction in the total gravity as a result of the removal of a larger amount of gas.  For $\fg=0.07$ and $c=11$ (not shown), the value of  $\sigma_u $  reaches $14\kms$ at $R\approx 2$. 

\begin{figure}
 \includegraphics[width=0.48\textwidth]{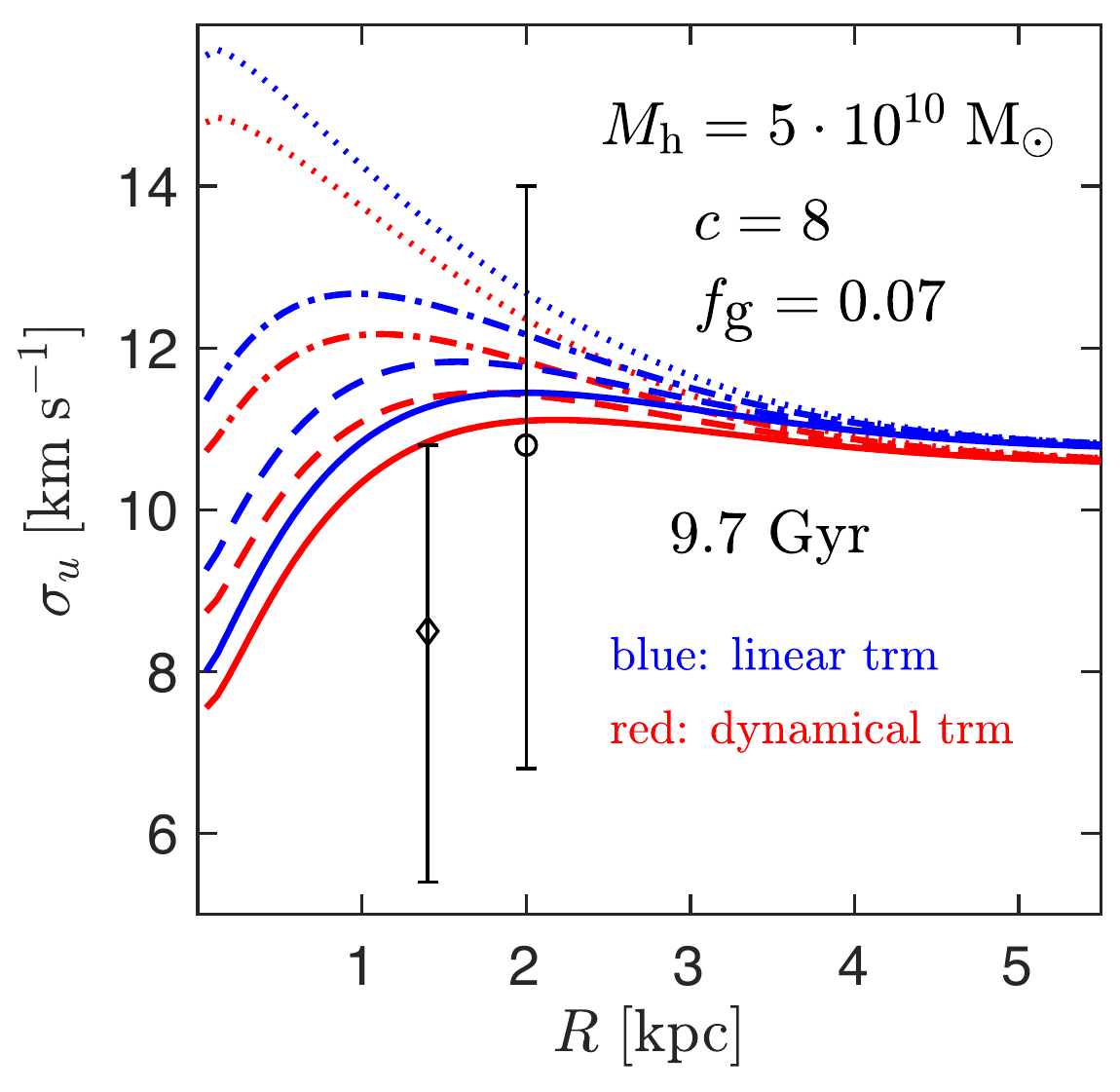}
  \vskip 0.0in
 \caption{ A comparison of the velocity dispersion obtained with linear and dynamical trimming recipes with fast gas removal. The different line-styles refer to different values of $\beta$, as in Fig.~\ref{fig:twoC}}
\label{fig:sigTD78}
\end{figure}

So far, we have considered the linear trimming recipe  III.a in \S\ref{sec:nscheme}. It is prudent to explore whether this recipe mimics, at least approximately, 
a realistic   stripping by the gravitational fields of a host halo. For this purpose, we apply the dynamical recipe III.b assuming the satellite moves on a circular orbit of radius 
 $300$ kpc in the gravitational field of an NFW host halo of virial mass $2.9\times 10^{12} \; \textrm{M}_\odot $ and $c=8$.  We run the simulation for the large-mass 
  case  $\Mv=5\times 10^{10}  \; \textrm{M}_\odot $ with $c=8$, where the equations of motion are written and numerically solved in the non-inertial frame attached to the satellite. As we already  have found for the linear trimming recipes, the corresponding dynamical effects are more pronounced for 
  the low gas fraction $\fg=0.07$. Hence, the simulation with dynamical trimming is run with this  gas fraction. 
   Curves of $\sigma_u$ from the simulation output at $9.7$ Gyr are plotted in red in Fig.~\ref{fig:sigTD78}. Also plotted, in blue, are the results from the linear trimming obtained previously (see Fig.~\ref{fig:bigM}).  It is reassuring that the two recipes yield similar results but perhaps not surprising since, in both cases,  the stripped matter is well outside $R\mathrm{e}$. Given the approximate nature of the analysis here, we do not make any further attempt to tune  the parameters of the host halo in order to bring the dynamical trimming even close to the linear trimming. The agreement seen in the figure is quite satisfactory for our purposes.

\subsection{High $z$ Halo}
\begin{figure} 
  \includegraphics[width=.48\textwidth]{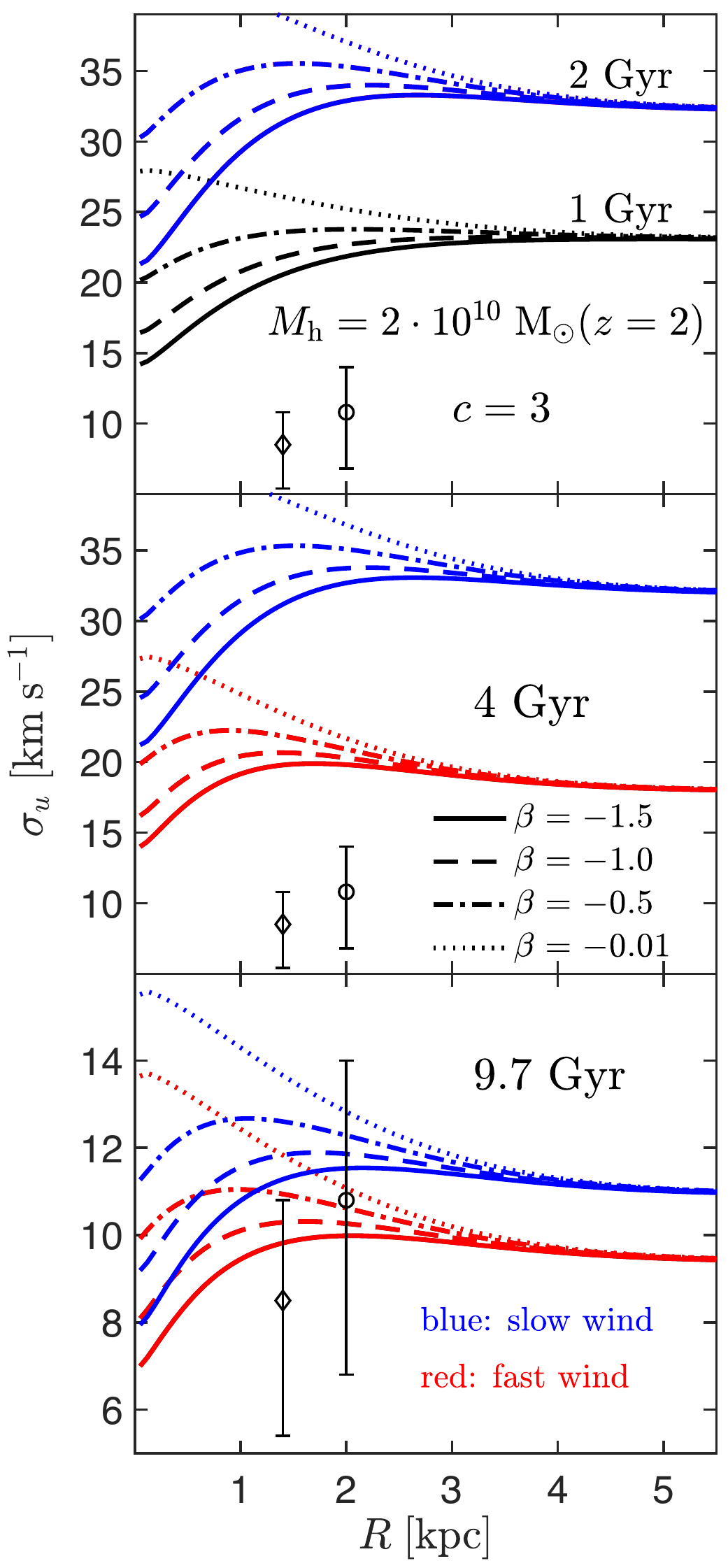} 
  \vskip 0.0in
 \caption{ The line of sight velocity dispersion within a projected distance $R$  from the simulations with  $\Mv=2\times 10^{10}\; \textrm{M}_\odot$   at $z=2$ 
 including linear trimming at $10$ kpc. The value $\fg=0.155$ is assumed.}
\label{fig:ztwo}
\end{figure}

\begin{figure} 
  \includegraphics[width=.48\textwidth]{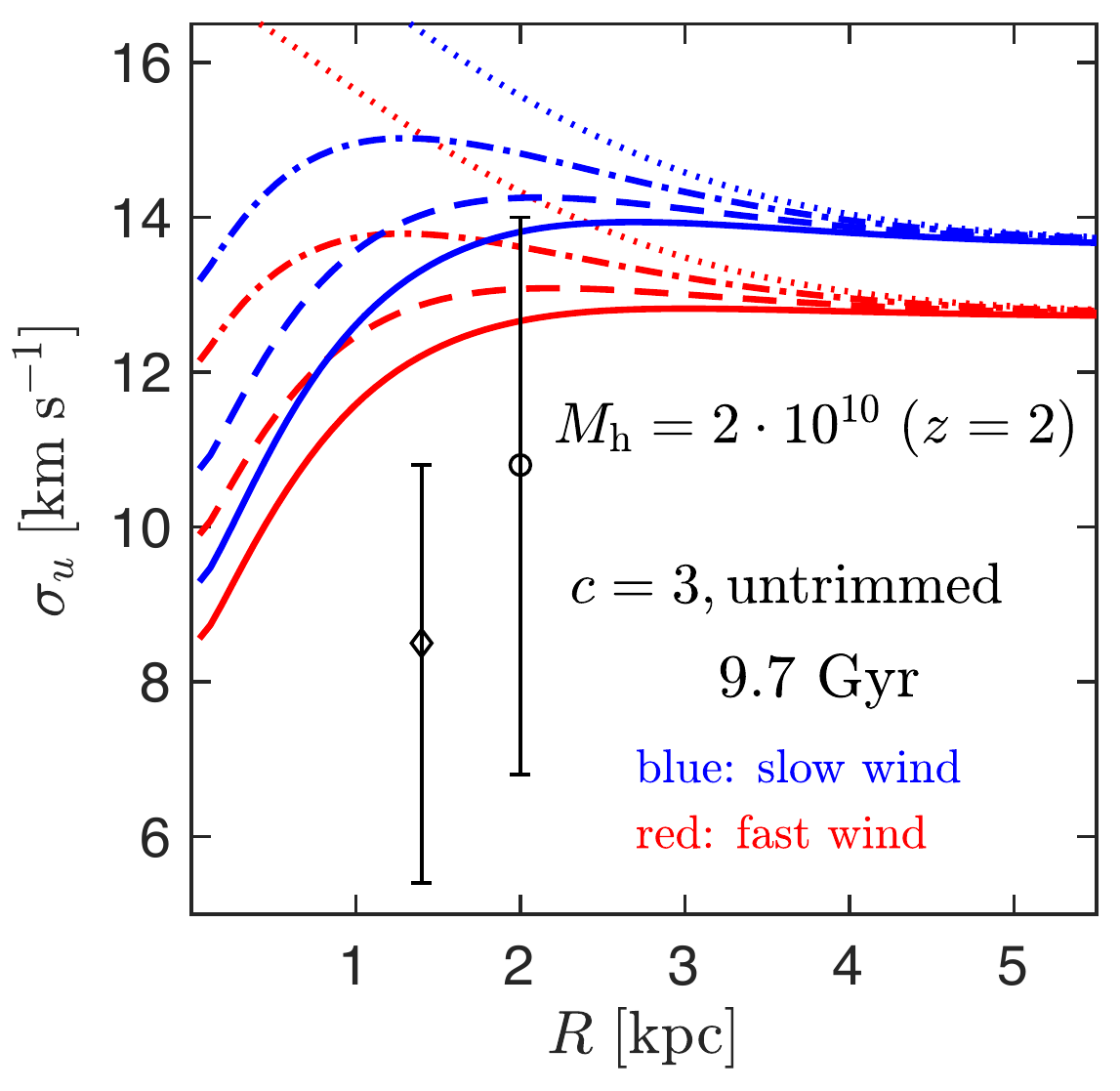} 
  \vskip 0.0in
 \caption{ The same as the bottom panel in the previous figure but without trimming of the halo.}
\label{fig:ztwountrm}
\end{figure}

{
Fig.~\ref{fig:ztwo} plots  the results for the high $z$  $\Mv=2\times 10^{10}\; \textrm{M}_\odot$   halo
with linear trimming,  in the same format as 
Fig.~\ref{fig:twoC}. 
Since the mass of the $z=2$ halo matches the median mass of the 
progenitor of the $z=0$ large $\Mv=5\times 10^{10}\; \textrm{M}_\odot$ halo, 
 we compare these results with those of Figure
\ref{fig:bigM} for the larger halos.
To match the amount of cool gas in the  two halos, a  fraction $\fg=0.07 $ for the $z=0$ high-mass halo should  be compared with $\fg=0.155$ for the lower-mass higher $z$ halo. }
{Before the initiation of the baryonic processes at 1 Gyr, we obtain similar 
$\sigma_u$ in the high-redshift halo and the $z=0$ large $\Mv=5\times 10^{10}\; \textrm{M}_\odot$ halo. 
Also,  at the   2 Gyr when cooling ends and at the final tine, 9.7 Gyr,  the corresponding curves in the two  halos agree well.  }
{Fast wind is more efficient at reducing $\sigma_u$. But both modes, fast and slow,  agree with the data at a reasonable level. 
Trimming is important for the high-$ z$ halo,  as seen in Fig.~\ref{fig:ztwountrm}, but 
$\sigma_u$ is still consistent with the \cite{Emsellem2018} measurement for the lower $\beta$. }

\subsection{Velocity dispersion at larger distances: The globular clusters}
\begin{figure}
 \includegraphics[width=0.48\textwidth]{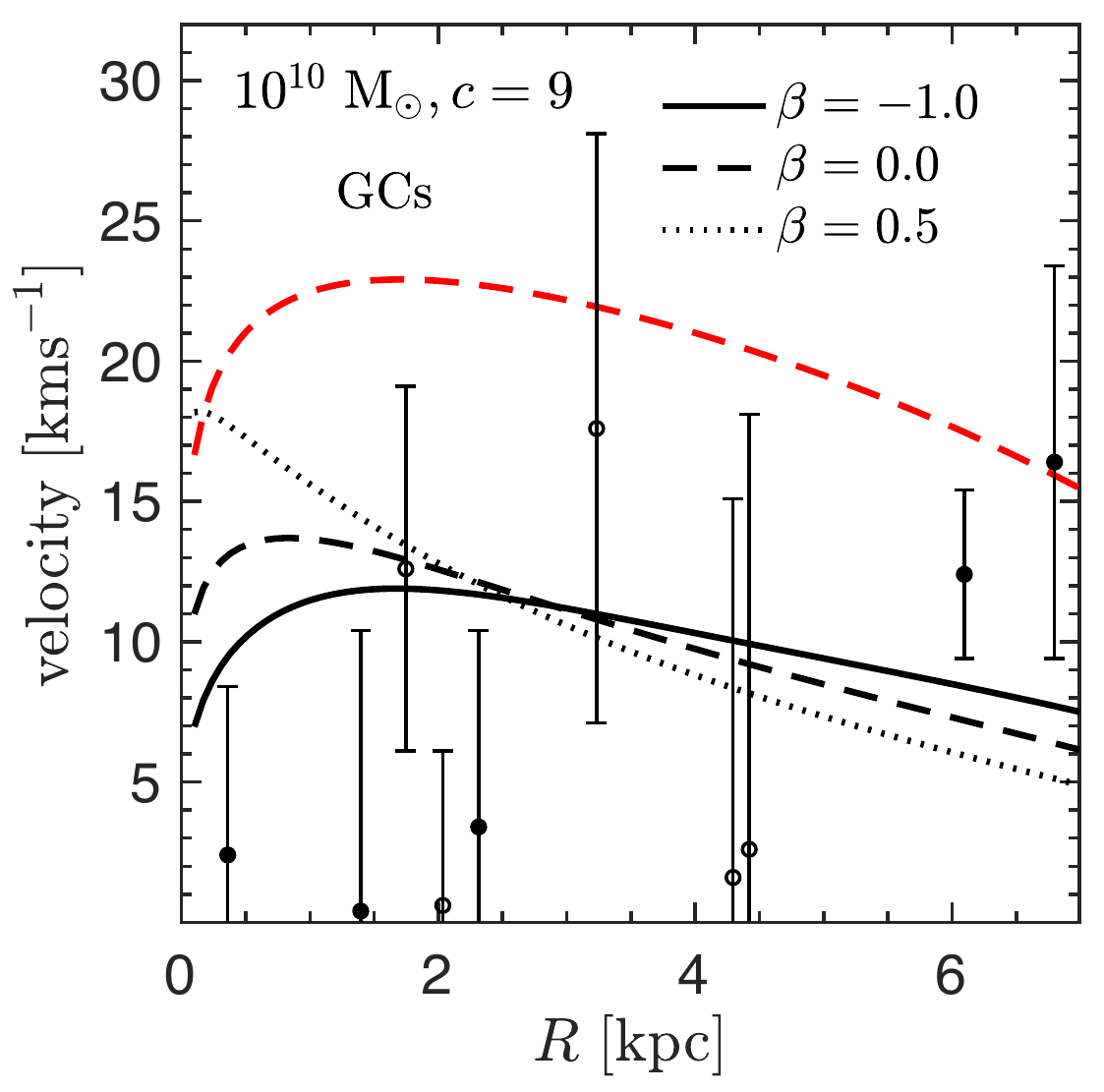} 
  \vskip 0.0in
 \caption{ The LOS measured velocities of GCs versus $R$ compared to the predicted $\sigma_u$ at $ R$ (rather than the average within $R$ as in all previous figures). Filled and open circles refer to positive (relative to the mean) and negative GC LOS. velocities.
 The red curve is derived from the initial NFW profile with mass and concentration as indicated in the figured. The black curves are for the final output of the simulation 
 with fast winds and no trimming for $\fg=0.155$. }
\label{fig:siguGC}
\end{figure}

We briefly discuss the agreement between the scenario presented here and the kinematical observations of GCs in \NDF . 
For these tracers, individual velocity measurements are available \citep{vanDokkum2018}. Unfortunately, the number of these tracers is small and the errors are large.  \cite{Nusser2019} presented an analysis based on the full phase space  DF under the assumption of tidal stripping, but without the inclusion of the effects of  gas cooling/ejection. We do not intend to apply the DF analysis to the model of the current paper. Instead,  we  provide a prediction for $\sigma_u$ and contrast  it with the GC measured velocities. This is done in Fig.~\ref{fig:siguGC} plotting $\sigma_u$ obtained  as described in \S\ref{sec:Jeans}, but with $n\propto r^{-2.3}$ which is  consistent with 
the observed distribution of the GCs on the sky \citep{Nusser2019, Trujillo2019}. 
The results are shown for four values of $\beta$, as indicated in the figure. Note that here we show  predictions for $\beta=0.5$, since a steady state DF of the form 
$L^{-2\beta}f_1(E)$ with $\beta=0.5$ can be found.  This is in contrast with the stellar distribution whose form 
did not allow for a DF of that form with a   nonnegative  $\beta$. We do not perform a full statistical analysis,  but it is evident that the curves of all plotted values of $\beta$ 
 are in reasonable agreement with the data. 

\section{Summary and conclusions}
\label{sec:con}

Independent of how ultra-diffuse galaxies form  \citep{Dalcanton1997,AL16,DiCintio2017,2017MNRAS.470.4231R}, we have outlined steps through which a halo with initial $\vc\ltsim 100 \kms$  can harbor a stellar component with  an observed velocity dispersion $\sigma\ll \vc$. The proposed scenario relies on common assumptions regarding 
galaxy formation that, when combined together, can lead to galaxies with kinematics similar to \NDF . The main ingredients are as follows: 1) an initially gas-rich satellite  galaxy with  a progenitor of a  halo of mass $ \ltsim 5\times 10^{10}\; \textrm{M}_\odot$ at $z=0$; 2)
an initial halo described by an NFW profile with a low concentration ($\gtsim 1\sigma$ deviation below the mean $c\!-\!M$ relation); 3) removal of a significant gas fraction either by  SN feedback or/and ram pressure; and 4) stripping of the outer parts of the halo  by the external tidal field. 

{
The scenario requires  the presence of  an initially  large cool-gas fraction.
This is a reasonable  assumption on account of the short cooling time scales. 
Indeed,  for optically selected local galaxies with  $5\ltsim \Ms/10^7 \textrm{M}_\odot\ltsim 100 $,   \cite{Papastergis2012} find  \hi\ mass 
 which could easily reach a factor of 10 larger than $\Ms$. The fraction of cool gas at high redshifts, before the bulk of the stars has formed, could be even  higher  in some galaxies. Measuring the neutral hydrogen fraction is very challenging at high redshift, but the observed large  fractions of molecular hydrogen in galaxies  at $z\sim 2$ \citep{Tacconi2010} implies a large amount of total cool gas as well.  }

{An initially   compact stellar component could  be puffed-up by the shallowing of the potential well by gas removal. The net outcome would 
be a diffuse stellar component with very low surface brightness \citep[e.g.][]{DiCintio2017}. The current work treated the  stellar  component in terms of  a fixed density profile and did not explore this possibility.    However, the details of this process in the context of low-velocity-dispersion galaxies,   could be tested with simulations, also  under the assumption of spherical symmetry, where the stellar component is treated as ``live" collisionless particles following a suitable density profile.  We leave that for a further study, where the 
sensitivity to the assumed initial stellar distribution is explored in detail. }

It is possible to envisage  the formation of completely DM free galaxies through galaxy collisions \citep{Silk2019} 
or though the condensation in gaseous  debris of galaxy mergers, as in the formation of tidal dwarf galaxies  \cite[e.g.][]{Lelli2015}.  Both possibilities seem to  require fine tuning in order to produce the properties of a galaxy like \NDF . 
In the current paper, we argue that there is  nothing  grossly unusual in the dynamics of  this galaxy and that it is like many other  small  galaxies  that show 
weak  evidence for dark matter.  We argue that the low velocity dispersion of  \NDF\  should not serve as an argument for a total lack of dark matter. The \NDF\ case  requires a combination of  additional ingredients such as a lower concentration (but still consistent with simulations) and tidal stripping, but all of these are natural in the 
standard cosmological paradigm. 
 
According to cosmological simulation, low concentrations are generally associated with more extended galaxies according to   the relation \citep{2019MNRAS.488.4801J} 
\begin{equation}
\label{eq:Re}
r_\textrm{e}=0.02\left(\frac{10}{c}\right)^{0.7}\rv\; , 
\end{equation}
where $r_\textrm{e}$  is the stellar half-mass radius. Taking $\rv=78$ kpc ($\Mv=5\times 10^{10}\; \textrm{M}_\odot$), the relation 
implies $1.45$ kpc for $c=11$, i.e. the mean $c$ for that mass \citep{Ludlow2016a} and 1.8 kpc for $c=8$. Taking $c=6$, which is  about a $2\sigma$ deviation from the mean $c\! -\! M$ relation \citep{Hellwing2016b}, we get $r_\textrm{e}=2.23$, compared to the observed $r_\textrm{e}\approx 1.3 R_\textrm{e}=2.6 $ kpc. 
Thus, the lower concentration may also help  increase $r_\textrm{e}$ as desired for UDGs.

We have seen that tidal stripping is helps lower  the velocity dispersion, but 
but, in contrast to \cite{Ogiya2018} and \cite{Carelton2018}, we do not require an elongated orbit of the satellite inside the host.  Instead of bringing the satellite close to the center of the host, we invoke SN feedback and ram pressure.   As argued in \cite{Nusser2019}, the kinematics of the relative speed between 
\NDF\ and the assumed parent galaxy NGC 1052 and the distance between them are likely hard to reconcile with elongated orbits for \NDF .

In \cite{Nusser2018}, the author  showed that orbital decay by dynamical  friction of some of the GCs in \NDF\  is expected.  
The conclusion was based on simulations that included a dark halo. 
\cite{DuttaChowdhury2019}  have studied the dynamical friction in this system for  the case of no dark matter at all. 
The presence of a core  in the 3D stellar distribution could suppress dynamical friction and even cause buoyancy \citep[e.g.][]{Cole2012}.
 In addition, scattering  among GCs themselves helps keep the GCs afloat around the core  at $r\sim 0.3R_\textrm{e}$. 
{ A  shortcoming of the analysis in \cite{DuttaChowdhury2019}  is that the GCs are treated as 
 point masses, and therefore, they cannot address the disruptive effects of GC-GC scattering on the GCs themselves. Nonetheless, these authors confirm that dynamical friction is also important (e.g.  figures 9 and 10 in their paper). They argue that the observations can be reconciled by 
 starting with a more spatially extended distribution of the GCs, but they do not offer
  any details of  a   realistic physical scenario for that setup. 
 The proposal of starting with a more extended GC distribution was also  noted  in \cite{Nusser2018} as possible solution. The point was originally  made by \cite{Angus2009} as a potential solution to the Fornax dynamical friction conundrum  \citep[see also][]{Boldrini2019}. A detailed physical formation scenario for 
 this type of initial conditions  remains lacking 
 \citep[but see][]{2020MNRAS.493..320L}. Thus, despite their conclusion regarding the consistency of the 
 dynamical friction argument with observations, their actual  findings actually agree with \cite{Nusser2018}.}
 
 {The scenario presented here requires strong velocity anisotropies with $\beta \ltsim -1$, with clear preference to nearly circular orbits of the stars.
 This may seem extreme, but the values invoked here are actually  favored in several cases \citep[e.g.][]{2019ApJ...874...41C}.
 More related to the current paper, \cite{2019ApJ...880...91V} find $\beta\approx -1$ from the kinematics of the UDG Dragonfly 44 under the assumption of NFW profile. }

{The scenario presented here offers, in  principle,  a way to prolong dynamical friction timescales. Outward movement of dark matter (of about a few $\kms$ in the 
inner few kpc and $\sim 10 \kms$ between $5$ and $10$ kpc) in response to the reduced gravity due to gas removal could actually result in a reduction in the net sinking rate of GCs to the center. 
However, a proper quantification of this effect is beyond the scope of the current paper. }
  
{We have  assumed spherical symmetry without any rotation.  Due to the low velocity dispersion of the system, the inclusion of even a mild rotational component  \citep{Emsellem2018} can also substantially boost the inferred mass estimate.
\citep[e.g.][]{Nusser2019,2020MNRAS.491L...1L}. }
\section*{Acknowledgements}
We thank the anonymous referee for useful comments. 
This research was supported by the I-CORE Program of the Planning and Budgeting Committee,
THE ISRAEL SCIENCE FOUNDATION (grants No. 1829/12 and No. 203/09). 
 \bibliography{Jeans.bbl}

 \end{document}